Emergence simulation of cell-like morphologies with evolutionary potential by virtual molecular interactions

Takeshi Ishida

Abstract
This study explores the emergence of life through a simulation model approach. The model "Multi-set chemical lattice model" is a model that allows virtual molecules of multiple types to be placed in each lattice cell on a two-dimensional space. This model is capable of describing a wide variety of states and interactions in a limited number of lattice cell spaces, such as diffusion, chemical reaction, and polymerization of virtual molecules. This model is also capable of describing a wide variety of states and interactions even in the limited lattice cell space of 100 x 100 cells. Furthermore it was considered energy metabolism and energy resources environment. It was able to reproduce the "evolution" in which a certain cell-like shapes adapted to the environment survives under conditions of decreasing amounts of energy resources in the environment. This enabled the emergence of cell-like shapes with the four minimum cellular requirements: boundary, metabolism, replication, and evolution, based solely on the interaction of virtual molecules.

1. Background

It is one of the great challenges of science to clarify the emergence process of the first life. Elucidation of this process will have a significant impact on research in the life sciences or micro-nano sciences. Toward the goal of elucidating this process, research concerning the origin of life has long been conducted and many hypotheses such as RNA world hypothesis[1] have been published. [for example 2), 3)]

When considering the origin of life, it is necessary to consider the conditions to determine as the emergence of life. While the definition of life as a premise needs to be clarified, the debate about the definition of life varies. Avoiding discussion and literature review here, but the four conditions that are generally acceptable were considered : 1) bounded, 2) replicating, 4) metabolizing energy, and 4) able to inherit information and to evolve. But with any one of these conditions, it cannot be called life. To reach the level of life, these functions must be fulfilled simultaneously.

And the following are some of the research approaches to the first cell emergence process that satisfy these four conditions. Various approaches are being taken, such as studies examining various chemical evolutions[4-5], experimental synthesis of minimal cells (protocells)[6-7], computer-assisted artificial life studies[8-9], and theoretical studies of complex systems[10].

Each research approaches also provide little explanation of how the four conditions for life were established. In explanations of various hypotheses, they are vaguely explained as "a cell membrane-like capsule made from some material was created", "a network of chemical reactions occurred among the various substances within it", "the result of a long process of trial and error" or as having the function of life when they reach a "certain level of complexity". It is not always clear how a chemical process that simultaneously satisfies all four conditions emerged after the materials for life were in place. Even if the first RNA is born, little is known about the process by which it acquires boundaries, metabolism, and replication functions.

Thus, the mechanism by which a persistent and evolving cell-like form can emerge under conditions of molecular interaction alone remains a major mystery. If this can be elucidated, it will not only increase the possibility of elucidating the origin of life and synthesizing the smallest cells (protocells), but it will also be applicable to the emergent construction and mass production of various molecular machines at the micro and nano level.

This study explores the emergence of life through a simulation model approach. A number of similar research approaches have been used. In the RNA world hypothesis, which is currently the leading hypothesis, there are examples of computer models that examine the processes leading to the RNA world. Szilágyi et al.[11] investigated the robust coevolution of catalytically active, metabolically cooperating prebiotic RNA replicators using an RNA World model of the origin of life based on physically and chemically plausible first principles. Yin et al.[12] and others have similarly constructed computer models of the emergence of RNA worlds.

Other computational approaches include the following; 1) Probabilistic models or discrete models such as cellular automata or artificial chemistry, and 2) calculating protocells from molecular dynamics models. A recent example of artificial life is Lenia[13]. This is a model based on Life Games that spins out life-like movements, but does not necessarily simulate the emergence of life's four conditions. Schmickl[14] made a simple motion law for moving and interacting self-propelled particles leading to a self-structuring, self-reproducing and self-sustaining life-like system. The patterns emerging within this system resemble patterns found in living organisms.

Regarding research on process of first life by artificial chemistry, Kruszewski et al.[15] validated self-reproducing metabolisms with a minimalistic artificial chemistry based on a Turing-complete rewriting system called combinatory logic. Fellermann[16] performed a dissipative particle dynamics simulation, combining self-assembly processes with chemical reaction networks. Moreover, Hutton[17] constructed a replication model by the interaction of two-dimensional particle swarms.

Regarding Molecular dynamics (MD) is a modeling method for calculating molecule behavior that requires high computation to calculate the behavior of a small number of molecules, as it solves each individual molecule mechanically. This method can calculate only a limited time within a limited space, making it difficult to compute a single whole living cell. For this reason, reducing the number of degrees of freedom is necessary to reduce computational complexity. There are many reports on coarse-grained models of MD, including computational examples of the self-organizing structure of cell membranes [18,19], of cell membrane division [20], and simulations of nanoscale mechanisms for nanovesicles' budding and fission[21]. In addition, stochastic-deterministic simulations over a cell cycle using a whole cell fully dynamical kinetic model have been reported[22]. These studies are simulations for each protocell construction process, but do not refer to the emergence of protocells from molecules and acquisition of the four conditions of life.

The author, Ishida, has been studying the physicochemical processes that simultaneously emerge under the four conditions of life from the aspect of discrete mathematical modeling. Ishida[23] improved the CA model, which is an external sum rule type CA , a model in which the state of a lattice is determined by the sum of the states of neighboring lattices, and found that various shapes such as self-replication, shown in Figure 1A , growth, and branching can be emerged and controlled with a few parameters. Furthermore, Ishida[24] considered a model that allows virtual molecules of multiple types to be placed in each cell on a two-dimensional space, as shown in Figure 1B. This model, referred to as the "Multi-set chemical lattice model", is capable of describing a wide variety of states and interactions in a limited number of lattice cell spaces, such as migration (diffusion), transformation (chemical reaction), and linkage (polymerization) of virtual molecules. This model is capable of describing a wide variety of states and interactions even in the limited lattice cell space of 100 x 100 cells.

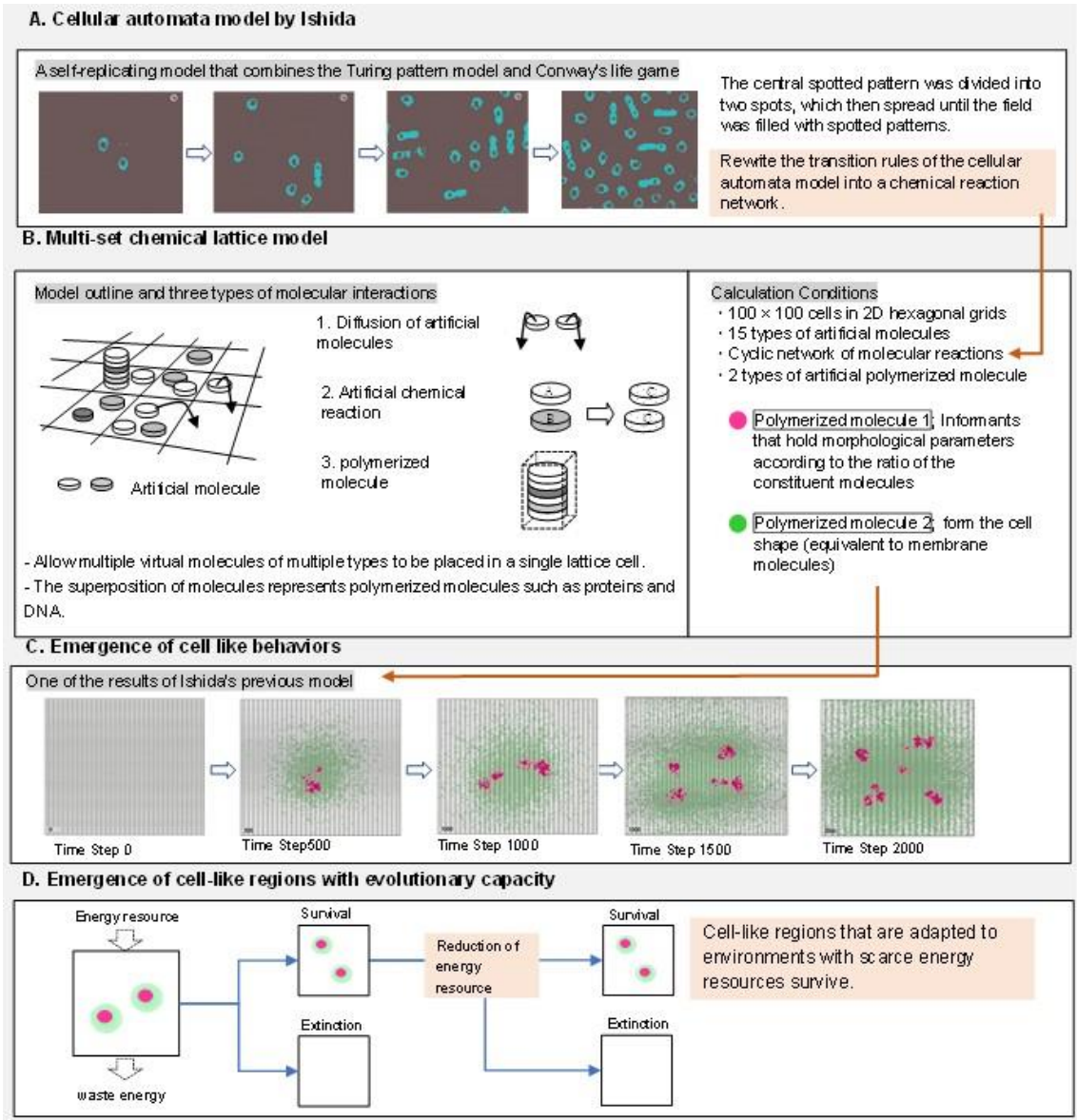

**Fig. 1. Schematic overview of the "Multi-set chemical lattice model".**  **A.** Ishida[23)] improved the cellular automata model and found that various shapes such as self-replication. **B.** This model allows multiple virtual molecules of multiple types to be placed in a single cell on two-dimensional lattice space. By applying discrete stochastic transitions to each molecule, describing a wide variety of states and interactions in a limited number of cells, such as migration (diffusion), transformation (chemical reaction), and linkage (polymerization) of virtual molecules. **C.** This figure shows the distribution of two types of polymerized molecules. In the initial stage of the time step, an accumulated region of red polymerized molecules was formed in the center of the lattice field, and then the region of polymerized molecules was continuously broken up. At the same time, it was confirmed that the information of the w value was retained in the green polymerized molecule and transmitted to the space. **D.** Constructing an extended model that explicitly incorporates energy metabolism, it was able to reproduce the evolution phenomenon in which a certain cell-like shapes adapted to the environment survives under conditions of decreasing amounts of energy resources in the environment.

In this Ishida's model[24], although 15 types of artificial molecules are assumed, the number of chemical reactions combinations among the 15 types molecules is enormous, and it is difficult to find the reactions among these combinations that cause life emergence phenomena. Ishida[24] converted the algorithm for generating Turing patterns identified in the author's papers 23 and 25 into the procedures for diffusion and reaction of 15 different virtual molecules and polymerization of virtual molecules. Using only the processes of molecular diffusion, reaction, and polymerization, shown in Figure 1C, this model was able to realize a process in which metabolism begins, boundaries are created, and these boundaries replicate, and retaining the information necessary for the maintenance of form (equivalent to genetic information).

However, this Ishida model[24] has been able to calculate the phenomena of simultaneous emergence of "boundary formation," "replication," and "metabolism," but was unable to specify whether the cell-like region retains evolutionary capacity. In this report, using this lattice-type multiset chemical model, it was constructed an extended model that explicitly incorporates energy metabolism. And as shown in Figure 1D, it was able to reproduce the phenomenon (i.e., this can be called "evolution") in which a certain cell-like shapes adapted to the environment survives under conditions of decreasing amounts of energy resources in the environment.

This enabled the emergence of a cell-like morphology with the four minimum cellular requirements: boundary, metabolism, replication, and evolution, based solely on the interaction of virtual molecules. This study neither provides a new hypothesis of life emergence nor determines a conventional hypothesis, but it is an important step in clarifying cellular emergence process.

In particular, the model assumes that the formation and decomposition of polymers consisting of 100 molecules linked together can be easily realized. However, this cannot be achieved in a natural reaction environment unless a catalyst is already present in biochemical reactions, and there are still difficulties in explaining the first life-emerging processes.

However, it proposes a concrete and simple mathematical model to explain the point that until now has only been vaguely expressed as "the conditions for life have been established as a result of a long process of trial and error". If it is possible to create a reaction system similar to the reaction process described in this paper through experimental approaches in the future, it will be possible to create cell-like capsules and mass-produce microcapsules.

## 2. Model
### 2-1 Overview of Multi-set chemical lattice model

2-1-1 Model Configuration

In this study, an extended two-dimensional lattice-type multiset chemical model by Ishida[24] was constructed. This Ishida model applies a multiset chemical model in lattice space. The computational representation of chemical reactions has been modeled as a multiset chemical model, which is also known as an "artificial chemical model"[26]. Multiset is the concept of a mathematical class plus multiplicity. Using the multiset concept, it is possible to construct a model that takes into account the number of molecules as well as the type of molecules. For example, considering molecule a and molecule b, a multiset representation of the state with three molecules a and two molecules b would be {a, a, a, b, b}. When these molecules transform molecule a and molecule b into molecule c through chemical reactions, it can be expressed by the following multiset rewriting rule.

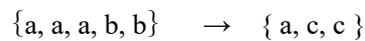

{a, a, a, b, b}   →   { a, c, c }

The rewriting rule can also be set to occur with a certain reaction probability. Once the rewriting rule and the reaction probability of the chemical reaction is determined, this modeling allows the number of molecules to be rewritten.

In each lattice cell, the number of molecules is recorded for each molecular species. The following is modeling method of the diffusion, reaction and polymerization processes of each molecule on the lattice cell.

As shown in Figure 2A, molecular diffusion can be represented by the exchange of molecular numbers between adjacent cells. In this model, the diffusion of molecules and polymerized molecules in each cell is represented by the following process, as shown in Figure 7A. The residual rate $r$ was defined here as an alternative parameter to the diffusion coefficient.

The residual ratio $r$ is the parameter of each molecular kinds, which is the fraction of unmoved molecules in each cell. The parameter $r$ is molecule-specific attribute values and fixed for all cells and all time-steps. As shown in Figure 7A, the following calculations are performed, where $b_{n,t}$ is the number of molecules in cell $n$ at time $t$: 1) A proportion, $b_{n,t} \times (1 - r) \times 1/6$, of the molecules diffuse toward the six adjacent cells evenly. 2) The residual, $b_{n,t} \times r$, molecules remain in the original cell. If the number of molecules is not an integer multiple of 6, the remainder of the molecules are distributed between adjacent cells with equal probability.

Next, as shown in Figure 2B, chemical reactions can be represented by rewriting the number of molecular types based on reaction probabilities, it's based on the idea of the multiset rewriting rule. In this model, 15 types of molecules were set up to replace the algorithm of previous studies of the Ishida model[23,25] with molecular reactions, as described below. Since this is an artificial chemical model, it does not correspond to real molecules, and molecular species are represented as "molecule 1", "molecule 2", .... The chemical reaction was then assumed to occur at each time step with a certain probability of change in the number of molecules in each cell, as shown in Figure 7B. For example, in the case of a reaction from molecule A to molecule B, molecule A is converted to molecule B with the reaction probability p. Assuming that the number of molecules A at time t is An and the number of molecules B is Bn, the number of molecules at the next time step is as follows.

$$A_{n,t+1} = A_{n,t} \times (1-p)$$
$$B_{n,t+1} = B_{n,t} + A_{n,t} \times p$$

Furthermore, macromolecules such as cell membranes and genes play an essential role in the emergent processes of life, and for this reason, methods to represent molecular polymers are necessary. In this model, It was considered a virtual box that is filled with virtual molecules collectively, as shown in Figure 7C. The empty box in the space was placed to wait the polymerization of molecules when the box was filled with molecules. When the box is emptied after removal of molecules from the box, it represents a state in which the polymerized molecules are broken down and returned to small molecules. Furthermore, the box itself was modeled to be diffuse in the lattice space.

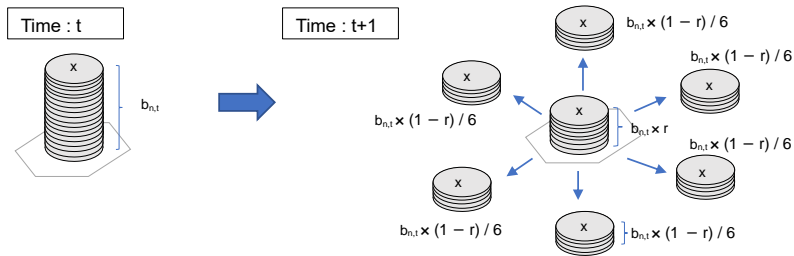
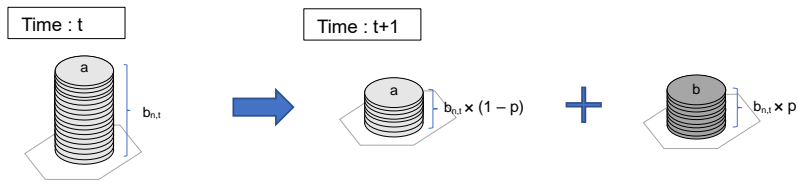
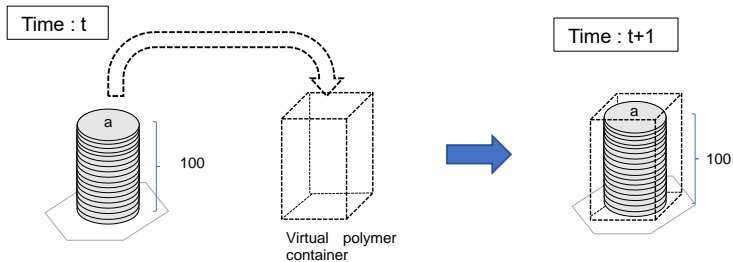

**Fig. 2. Schematic of molecules diffusion, chemical reaction and virtual box model to represent the**

**state of polymers.** **(A)** Molecular diffusion can be represented by the exchange of molecular numbers between adjacent cells. In this model, the diffusion of molecules and polymerized molecules in each cell is represented by the following process. The residual rate was defined here as an alternative parameter to the diffusion coefficient. The residual ratio r is the parameter of each molecular kinds, which is the fraction of unmoved molecules in each cell. The parameter r is a molecule-specific attribute value, fixed for all cells and all time-steps. As shown in Figure, the following calculations are performed, where $b_{n,t}$ is the number of molecules in cell n at time t: 1) A proportion, $b_{n,t} \times (1 - r) \times 1/6$, of the molecules diffuse toward the six adjacent cells evenly. 2) The residual, $b_{n,t} \times r$, molecules remain in the original cell. If the number of molecules is not an integer multiple of 6, the remainder of the molecules are distributed between adjacent cells with equal probability. **(B)** Chemical reactions is represented by rewriting the number of molecular types based on reaction probabilities. Figure shows one example of the case of a reaction from molecule a to molecule b with reaction rate p. **(C)** It was assumed a virtual box that is filled with virtual molecules collectively. The empty box in the lattice space was placed to represent the polymerization of molecules when the box was filled with molecules. When the box is emptied after removal of molecules from the box, it represents a state in which the polymerized molecules are broken down and returned to small molecules.

2-1-2 Cinfigulation of molecular reaction network

In the Ishida model[24] , 15 different molecules are set up to replace the algorithm of the Ishida model[23, 25] , which realized the emergence of cell-like shape in the cellular automaton model, with molecular reactions. The 15 molecules have the following roles.

- Molecule 1 ； Molecule that is the material to be converted into each molecule. (Initially, a large number of them are placed in the lattice space.)
- Molecule 2, Molecule 3; Molecules corresponding to diffusing substances in the Turing pattern model. (The difference of diffusion coefficients is expressed by the difference in their residual rate.)
- Molecule 4, Molecule 5; substances that change from Molecules 2 and 3 during diffusion, respectively.
- Molecule 6, Molecule 7; Molecules that are materials that constitute "polymerized molecule 1". Polymerized molecule 1 is a polymer of molecules 6 and 7, and the ratio of molecules 6 and 7 represents the morphology parameter w. Ishida model [26] algorithm, depending on the value of the morphological parameter, Turing pattern can be controlled from black spots on white, stripe patterns, and white spots on black .
- Molecules 8, 9, 10, 11; Molecules for describing the transition equation of the Ishida model[23] in chemical reactions.

- Molecule 12; Molecule that is the material that makes up the "polymerized molecule 2" that represents the boundary of the cell.
- Molecules 13, 14, 15; Molecules for describing the transition equation of the Ishida model[23] in chemical reactions.

And in the Ishida model [24], the whole reaction pathways for each molecule are set up as shown in Figure 3. By setting up the molecular diffusion and reaction in each lattice as described above, it is possible to realize algorithms similar to the Ishida models[23] and to generate a variety of forms including Turing patterns. The system then sets up a cyclical reaction path where molecules decompose at a certain ratio and return to molecule 1, and this chain of reactions is thought to be modeled metabolism at the same time.

Furthermore, based on this molecular setup in this study, it was added new molecules related to energy metabolism and their reactions. Specific descriptions are presented in the next section.

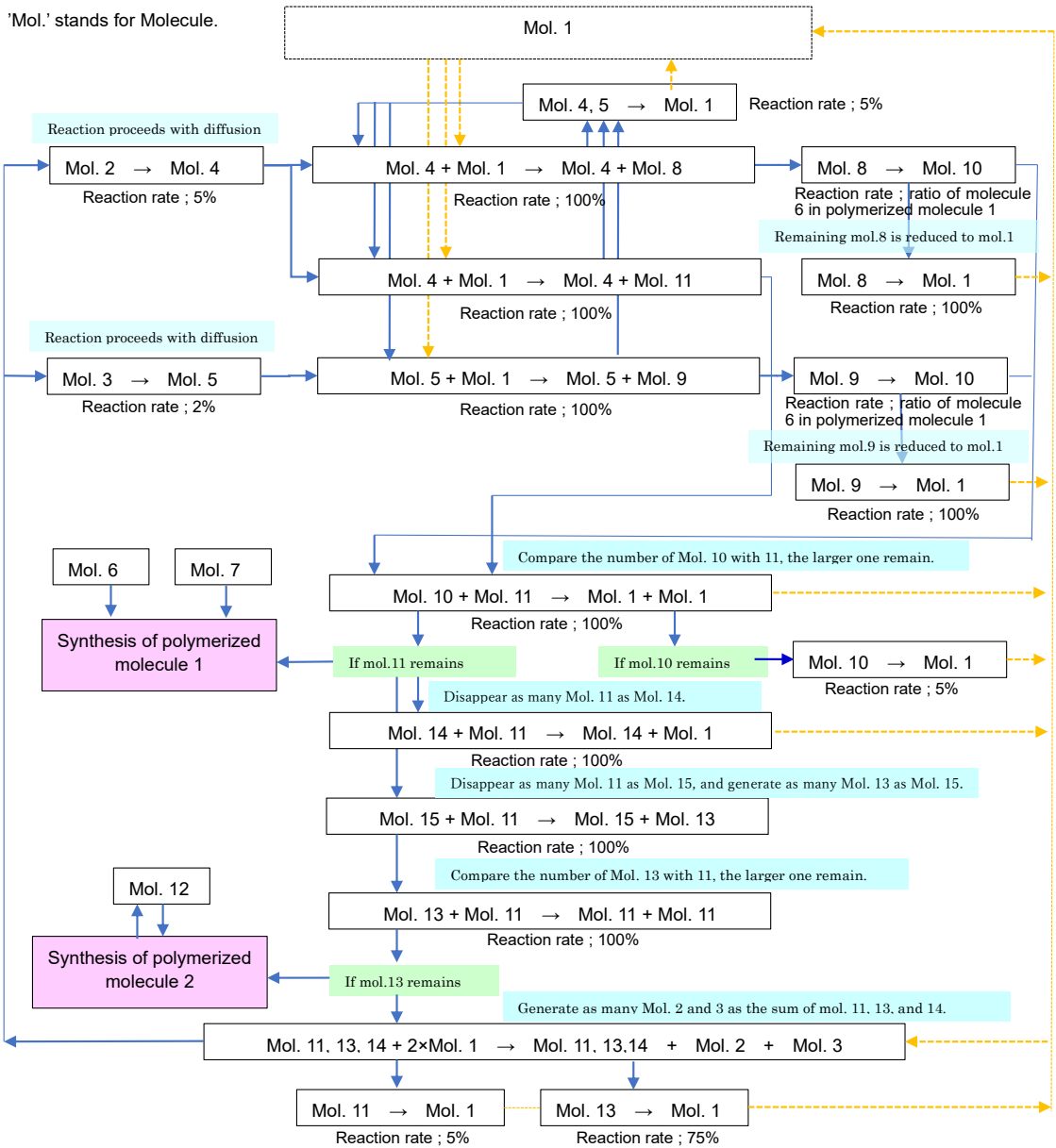

**Fig. 3. Chemical reactions network map**[24]**.** This figure summarizes the network of chemical reactions in Ishida model[24]. Molecular diffusion and a cyclic network of molecular reactions were constructed to create an algorithm identical to the Ishida model[23]. These reaction networks allows for the formation of cell-like shapes and the emergence of self-replication.

2-2 Introduction of Energy Metabolic Reactions
2-2-1 Supply and removal of energy resources to lattice space
The former Ishida model[24] is built on the assumption that there is enough energy in the lattice space for all chemical reactions to proceed, and does not explicitly incorporate energy metabolism. To this model in this study, the model was extended to allow the introduction of molecules that serve as energy resources and the molecules that are consumed as energy during each chemical reaction.

First, molecule 20 is introduced as energy resources, molecule 25 carries energy to assist the chemical reactions of other molecules, and molecule19 is waste molecules from which energy was taken from the 25 molecule. The resource molecule 20s are continuously supplied in constant amounts throughout the lattice space, while the waste molecule 19s are excluded from the space. These molecules allow for the explicit inflow and removal of energy into and out of the lattice space.

The reactions of the energy molecules were set up as follows.

Molecule 20 → Molecule 25

> Reaction conditions: There must be a polymerized molecule 1 with a molecular arrangement that serves as a catalyst.
>
> Reaction rate: It is set that the reaction rate varies depending on the amount of a specific catalyst sequence (the part where five molecule 6s are connected in succession) in the polymerized molecule 1.

Also, in each of the reactions in the reaction network shown in Figure 3, molecule 25 was set to be consumed with each reaction as a reaction energy.

Molecule 25 → Molecule 19 +energy

Reaction conditions: Molecule 25 is consumed during the reaction of each molecule according to the number of molecules and converted to molecule 19

Reaction rate: 100%.

From the above, in order to maintain the cell-like shape in which the polymerized molecules 2 are assembled, the molecules 25 must continue to be supplied to the area. In order for molecule 25s to exist, molecule 20s must be supplied to each lattice cell.

2-2-2 Catalytic function set up by polymerization molecule 1

In the reaction that converts molecule 20 to molecule 25, polymerized molecule 1 was assumed the catalyst. As shown in Figure 4a, polymerized molecule 1 is composed of molecule 6 and molecule 7 in a specific ratio, and the ratio is a parameter of morphology, but at the same time, a specific sequence of molecule 6 is assumed to have a catalytic function.

For the composition ratio of molecule 6 and molecule 7, 0.68 (molecule 6/(molecule 6 + molecule 7)) was used in this paper, referring to the Ishida model[24] setting. This value is a parameter that allows the emergence of cell-like shapes and the emergence of continuously dividing state in the previous model. In the previous model, only the ratio of these molecules was used as a parameter for morphogenesis, and the order of molecules 6 and 7 was not considered.

In this study, the ratio of molecule 6 to molecule 7 was used as a parameter, as well as a specific sequence of molecule 6 as a catalyst. Specifically, as shown in Figure 4B, it was assumed that if there are five consecutive sections of molecule 6s, this part has a catalytic function to produce 25 energy molecules. Such a setting is an analogy for a particular sequence of amino acids performing a particular catalytic function.

In this study, as shown in Figure 4C, the number of consecutive portions of molecule 6 in the polymerization molecule 1 in each lattice was counted, and the reaction probability was set to change according to the number of configurations of five consecutive portions of molecule 6. In this study, as shown in Figure 4C, the number of consecutive portions of molecule 6 in the polymerization molecule 1 in each lattice was counted, and the reaction probability was set to change according to the number of configurations of five consecutive portions of molecule 6. Although any sequence can be used as the catalytic function, this setting was chosen because it is easier to simulate the frequency of five consecutive sections when polymerized molecule 1 is randomly mutated.

The specific reaction probability formulas were set up as follows. If molecules 6 and 7 in polymerized molecule 1 mutate randomly, the composition ratio of 5 consecutive molecule 6s changes by approximately 0.15 to 0.3, so the reaction rate changes from 0% to 60%.

Reaction rate of energy molecule 20 = Composition ratio of 5 consecutive molecule 6s × 4.0 - 0.6

A. The ratio of molecule 6 in the polymerized molecule 1 holds the numerical value of the morphology parameter.

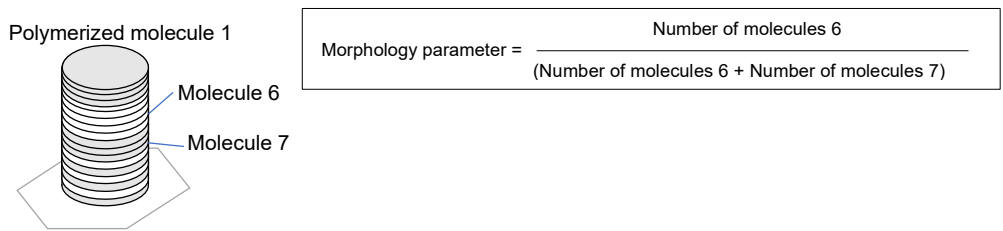

B. Polymerized molecule 1 is consists of molecule 6 and molecule 7.

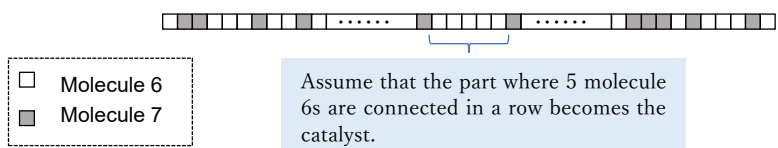

Assume that the part where 5 molecule 6s are connected in a row becomes the catalyst.

C. Energy production varies with the number of consecutive molecules 6 in the polymerized molecule 1.

$$\text{Ratio of five consecutive molecules 6} = \frac{A_5}{A_1 + A_2 + A_3 + A_4 + A_5}$$

$A_1$ = Number of solo molecules6

$A_2$ = Number of two consecutive molecules 6

$A_3$ = Number of three consecutive molecules 6

$A_4$ = Number of four consecutive molecules 6

$A_5$ = Number of five consecutive molecules 6

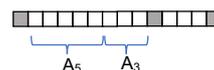

If there are 5 or more consecutive molecules 6, divide them into 5 or less

Reaction rate of energetic molecule = Ratio of five consecutive molecules 6 × 4.0 – 0.6
  (Mol. 20 --> Mol.25)

図 4　触媒のモデル化

**Fig. 4. Catalyst modeling. A.** Polymerized molecule 1 is composed of molecule 6 and molecule 7 in a specific ratio, and the ratio is a parameter of morphology. **B.** The ratio of molecule 6 to molecule 7 was used as a parameter, as well as a specific sequence of molecule 6 as a catalyst. Specifically, it was assumed that if there are five consecutive sections of molecule 6s, this part has a catalytic function to produce 25 energy molecules. **C.** The number of consecutive portions of molecule 6 in the polymerization molecule 1 in each lattice was counted, and the reaction probability was set to change according to the number of configurations of five consecutive portions of molecule 6.

2-2-3 Initial configuration of molecule 6 in polymerized molecule 1

　In the initial state of lattice space, if there are no polymerized molecules 1 at all, no energy molecules 25 will be produced and the reaction network will not proceed, so some polymerized molecules 1 were placed in the lattice space as an initial value. These polymerized molecules 1 were created and arranged so that the ratio of the degree of continuity of molecule 6 is in the configuration shown in the table in Figure 5A. This ratio is the average of the ratios at which each continuity appears when 100 molecules are randomly arranged in a 0.68:0.32 ratio for molecule 6 and molecule 7, and then randomly mutated (molecule 6 and molecule 7 are randomly swapped). This can be calculated by creating a simple Excel sheet.

The replication conditions of polymerized molecule 1 is the same as in the previous model, which conditions are the molecule 11 is present in the cell and there are sufficient numbers of molecules 6 and 7. In this model, it was introduced a setting in which a mutation occurs by the random exchange of molecules 6 and 7 with a certain probability.

2-4 Model implementation and calculation conditions

The model was programmed in JavaScript language.

2-4-1 Configuration of lattice cell grids

This model used 2D hexagonal grids wherein the transition rules are simple to apply. This models were implemented under the following conditions:
  - Calculation field: 100 × 100 cells in hexagonal grids
  - Periodic boundary condition

2-4-2 Conditions of calculations

The diffusion of molecules can be varied by the residual rate of each molecule, and the values in Table 1 were set as the standard case, referring to the settings in the previous model. The higher the residual rate, the higher the percentage of molecules remaining in the lattice cell. The smaller the value, the more quickly diffused.

2-4-3 Initial conditions

As an initial condition, molecule 1 was placed throughout the lattice with 1,000,000 molecules in each cell, molecule 6 and 7 with 50,000 molecules, and molecule 12 with 100,000 molecules. At the same time, 100 energy resource molecule 20s and 100 energy molecule 25s were placed.

In addition, place 100 molecules of Molecule 2 and 3 in some cells in the center area of lattice field (Fig. 5B). In addition, ten polymerized molecule 1s, which was polymerized with ratio molecules 6 and 7 with morphological parameter w, were placed in the central region of the space as shown in Fig. 5C.

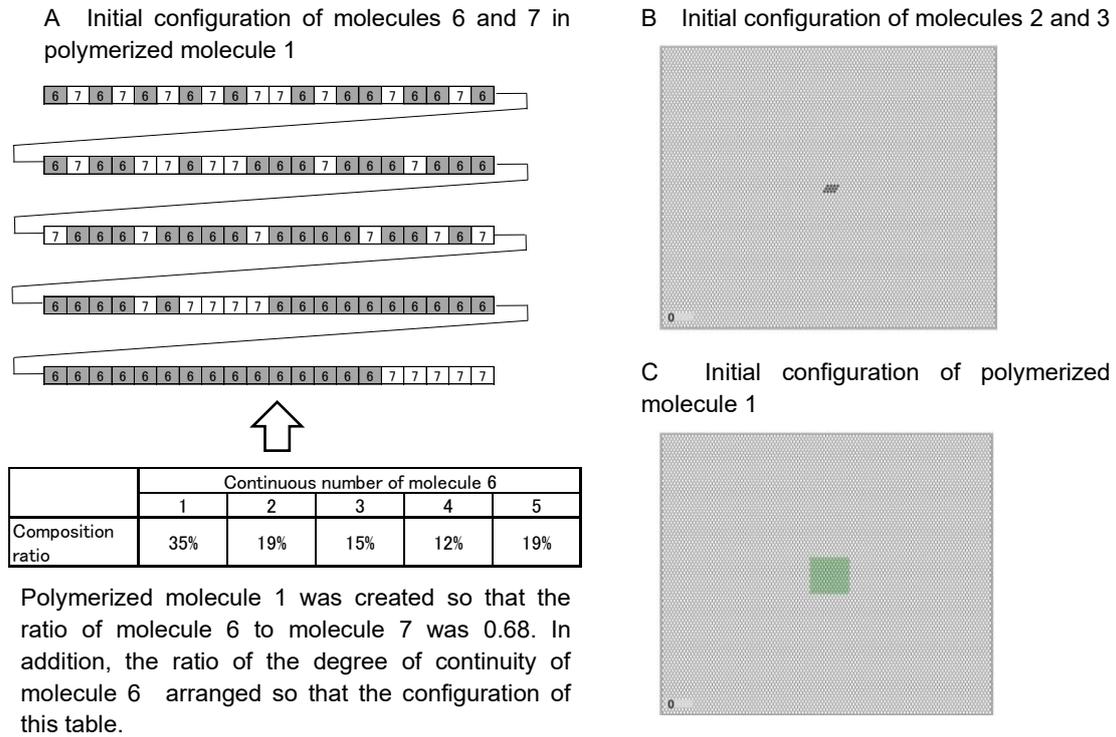

**Fig. 5. Initial arrangement of molecules in polymerization molecule 1 (catalyst) and initial arrangement of molecule2 and3. A.** Polymerized molecules 1 was created and arranged so that the ratio of the degree of continuity of molecule 6 is in the configuration shown in the table in Figure 5A. **B.** Placed 100 molecules of Molecule 2 and 3 in the center area of lattice cell field. **C.** Ten polymerized molecule 1s, which were polymerized with ratio molecules 6 and 7 with morphological parameter w, were placed in the central region of the space.

For the determination of the standard case value of each parameter, the parameters were adjusted through several trials and errors, referring the values of the previous Ishida model. Table 1 shows the values of each parameter for the standard case.

Table 1: Standard parameters set and the settings for the calculation cases with modified parameters for the standard case.

|  | Residual rate with diffusion process | Initial arranged number of entire computational space | Initial arranged number of central initial placement area |
|---|---|---|---|
| Molecule 1 | 0.0 | 1,000,000 |  |
| Molecule 2 | 0.75 |  | 100 |
| Molecule 3 | 0.05 |  | 100 |
| Molecule 4 | = Molecule 2 |  |  |
| Molecule 5 | = Molecule 4 |  |  |
| Molecule 6 | 0.0 | 50,000 |  |
| Molecule 7 | 0.0 | 50,000 |  |
| Molecule 8 | 1.0 |  |  |
| Molecule 9 | 1.0 |  |  |
| Molecule 10 | 1.0 |  |  |
| Molecule 11 | 1.0 |  |  |
| Molecule 12 | 0.0 | 100,000 |  |
| Molecule 13 | 1.0 |  |  |
| Molecule 14 | 1.0 | 3 |  |
| Molecule 15 | 1.0 | 9 |  |
| Molecule 19 | 0.0 |  |  |
| Molecule 20 | 0.4 | 100 | 1,000 |
| Molecule 25 | 0.2 | 100 | 500 |
| Polymerized molecule 1 | 0.75 | 0 | 10 |
| Polymerized molecule 2 | 0.75 | 0 |  |

2-4-4 Calculation case

The effects of the major parameters on the calculation results are summarized in the Ishida model[24]. The parameters investigated in this study are the amount of energy resources E (the number of molecule 20s) supplied to the lattice space at each computational step and the mutation rate of the polymerized molecule 1. In each cell, it was assumed that the following number of molecule 20s would be supplied in the next step if molecule 20 was zero.

$$E = 200, 300, 400, 500, 600, 700, 800$$

In order to simulate an environment in which energy resources decrease with time, a setting in which energy resources E are reduced after a certain time step was also calculated at the same time. The number of resources was reduced to 200 after the 4000 step and to 100 after the 6000 step, relative to the initial supply of resources. Furthermore, the mutation rate was calculated for 0.1, 0.2, 0.3, 0.4, and 0.5, using 0.1 as the standard.

3. Results

3-1 Simulation of cell-like shapes under conditions of abundant supply of energy resources

The results of the time series of the case where the amount of energy resources E input into the computational space is 800 are shown in Figure 6A. In the figure, polymerized molecule 2, which is responsible for cell morphology, is shown in red. Under conditions of abundant energy resources, this cell-like shape emerges and divides, reproducing the same phenomenon as in Ishida's previous model.

Even if the initial conditions and parameter values are the same, the pattern of cell-like patterns changes with each calculation. This is due to the fact that excess molecules that cannot be divided into the six orientations in the molecular diffusion process are assigned to the orientations by random numbers, and that there are parts where the presence or absence of molecular conversion is determined probabilistically based on reaction probabilities.

Figure 6B shows the amount of polymerized molecule 1 (green area in the figure), which plays the role of an informant, generated and degraded. It can be seen that as the cell-like regions divide and increase in number within the computational domain, the amount of polymerized molecule 1 (informant) produced and degraded also increases. Thus, informants are repeatedly generated and decomposed, and by introducing the action of selection through mutation here, it is possible to change the information held by informants.

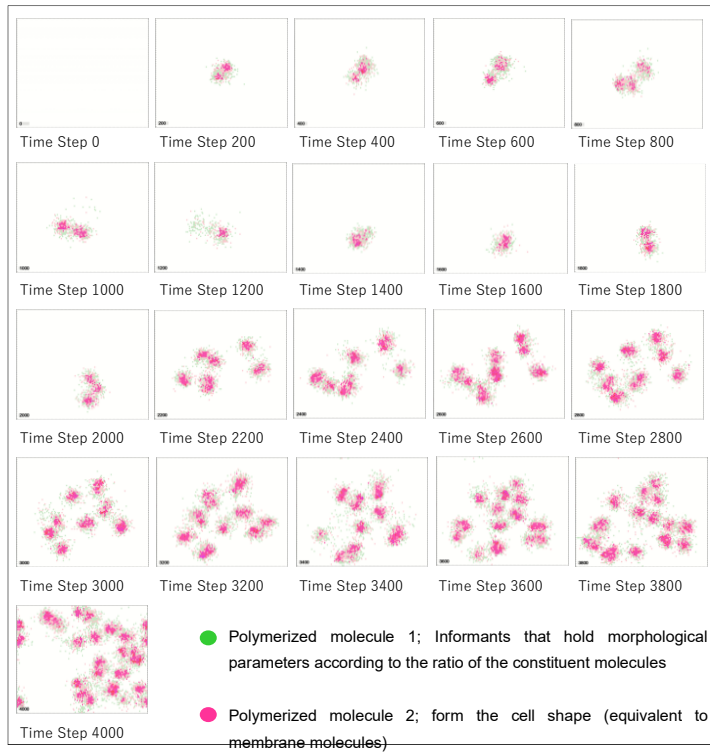
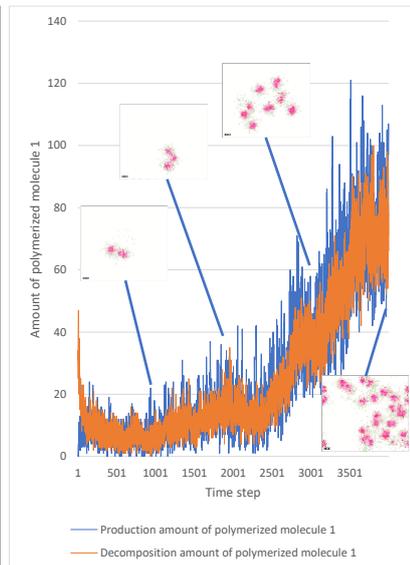

**Fig. 6. The results of the time series of the case where the amount of energy resources E input into the computational space is 800. A.** Polymerized molecule 2, which is responsible for cell morphology, is shown in red. Under conditions of abundant energy resources, this cell-like shape emerges and divides. **B.** Amount of polymerized molecule 1 (green area in the figure), which plays the role of an informant, generated and degraded. It can be seen that as the cell-like regions divide and increase in number within the computational domain, the amount of polymerized molecule 1 (informant) produced and degraded also increases..

3-2 Results with energy resource supply

The results of 10 calculations for varying the amount of energy resource E (the amount of new supply when the number of molecule 20 is zero in each lattice cell) are shown in Figure 7A. The vertical axis in Figure 7A shows the evolution of the ratio of the number of five consecutive molecule 6s in the polymerized molecule. As mentioned above, when molecules 6 and 7 in polymerized molecule 1 are randomly exchanged, five consecutive molecules 6s have an average occurrence rate of 19% when the composition ratio of molecule 6 is 0.68. All of the surviving computational cases in the 10 trials had an occurrence rate of 19% or higher, indicating that cell-like regions with a high occurrence rate of polymerized molecule 1 survived in the computational lattice space. Cell-like regions with informants that have a larger ratio of five consecutive molecule 6s are more likely to survive by producing more energy molecule 25s. When the energy resource supply E is lower, the amount of energy resource input into the lattice space is reduced and the number of cases where the cell shape disappears in a short time step increases.

Figure 7B shows the results of 10 trials for each energy resource E, showing the number of survived times and disappeared times. When the resource supply E exceeded 500, the number of cases surviving increased. It can be seen that the greater the supply of energy resource E, the more often the cell-like regions tend to survive.

3-4 Results with mutation rates of polymerized molecule 1

In the standard case, the mutation of polymerized molecule 1 is set to 0.1. Figure 7C shows the results when the mutation rate is varied. This figure shows the number of survivors and the number of extinctions for each of 10 trials with different mutation rates in the case of energy resource E = 500. The results show that higher mutation rates do not increase the number of cell-like regions that survive. It is thought that this is because a high mutation rate increases the frequency of replacement of molecule 6 and molecule 7, making it difficult to maintain five consecutive portions of molecule 6s for a long period of time, and thus the survival ratio cannot be kept high.

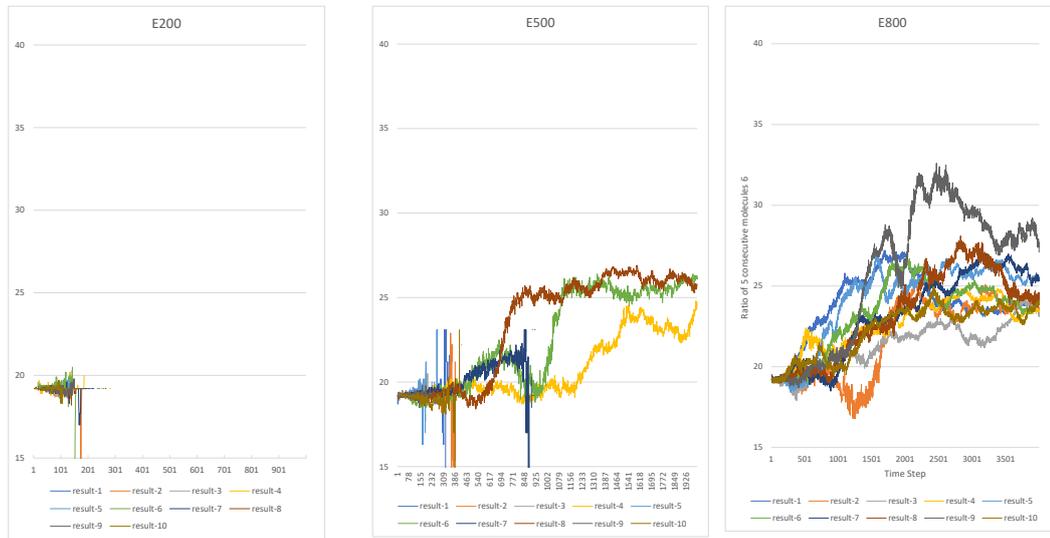

A. Ratio of 5 consecutive molecular 6 (Energy resource =200, 500, 800, Mutation rate = 0.1)

B. Number of survivors/extinctions with the amount of energy resources

C. Number of survivors/extinctions with the mutation rate

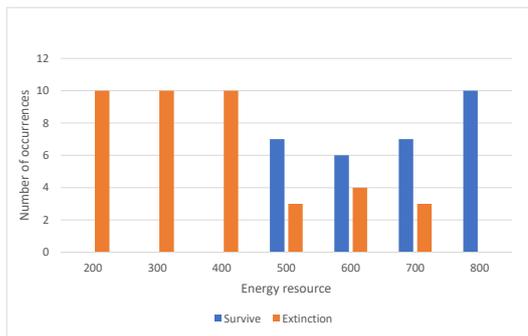
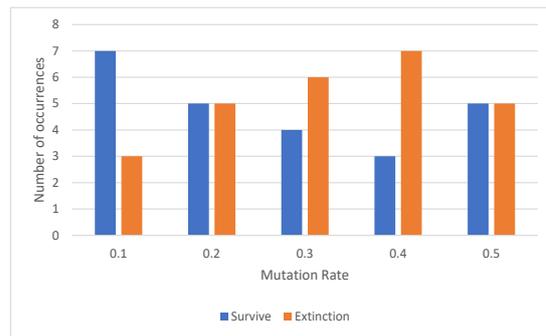

**Fig. 7. Calculation results for different amounts of energy resources and mutation rates. A.** These are the results of 10 calculations for varying the amount of energy resource E (the amount of new supply when the number of molecule 20 is zero in each lattice cell). The vertical axis in Figure 7A shows the evolution of the ratio of the number of five consecutive molecule 6 in the polymerized molecule. **B.** These are results of 10 trials for each energy resource E, showing the number of survived times and disappeared times. When the resource supply E exceeded 500, the number of cases surviving increased. It can be seen that the greater the supply of energy resource E, the more often the cell-like regions tend to survive. **C.** This is the results when the mutation rate is varied. This figure shows the number of survivors and the number of extinctions for each of 10 trials with different mutation rates in the case of energy resource E = 500.

3-5 Calculation results in an environment of decreasing energy resource supply

Figure 8 shows the results of reducing the amount of energy resources supplied to the lattice space over time. As in Figure 7A, the ratio of the number of occurrences of five consecutive molecule 6s in polymerized molecule 1 is shown. In the figure, the results of the calculations are shown for three cases, one in which the amount of energy resources E is continued at 800, another three cases in which the amount of energy resources is maintained at 800 until step 4000, reduced to E = 200 after step 4000, and reduced to E = 100 after step 6000.

This result indicates that cell-like regions will survive even when the amount of energy resources supplied to the lattice space is reduced. Figure 7B shows that the environment at E = 200 is difficult for cell-like regions to survive, but cell-like regions survived even when the amount of energy resources is 200 or even 100 after the 4000 step. This suggests that regions that were able to increase the number of consecutive sections of five molecule 6s were able to survive.

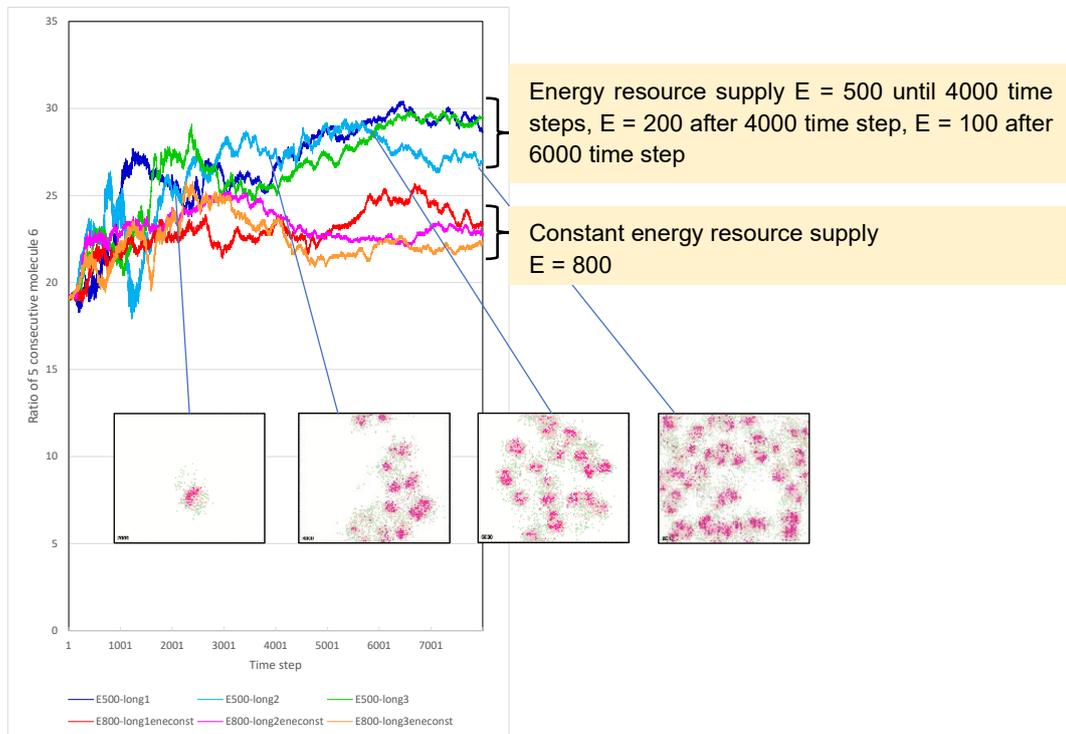

**Fig. 8. Calculation results in an environment of decreasing energy resource supply. A.** These are the results of reducing the amount of energy resources supplied to the grid space over time. As in Figure 7a, the ratio of the number of occurrences of five consecutive molecule 6s in polymerized molecule 1 is shown. In the figure, the results of the calculations are shown for three cases, one in which the amount of energy resources E is continued at 800, another three cases in which the amount of energy resources is maintained at 800 until step 4000, reduced to E = 200 after step 4000, and reduced to E = 100 after step 6000.

4. Discussion

In this model, 18 artificial molecules were assumed, and the unique residual rate of each molecule is set, as well as the reaction probability of intermolecular reactions and the molecular ratio of molecule 6 to molecule 7 of polymerized molecule 1 (representing the parameter of morphology). Molecules were placed in each cell as initial values, and local numerical manipulation of only diffusion, reaction, polymerization, and decomposition of each type of molecule allowed us to simulate the emergence and replication of cell-like shapes in which polymerized molecules2 aggregate. It was able to construct a model in which the information on the morphological parameter is retained in the form of the molecular configuration ratio of polymerized molecule 1, and this configuration ratio is inherited in the process of the emergence and disappearance of self-replicating regions.

Furthermore, it was introduced mutation process in which the composition ratio of molecules 6 and 7 of polymerized molecule 1 is not changed, but the order of the molecules is randomly changed, and it was also introduced a setting in which the amount of energy (molecule 25) production changes according to the number of five consecutive molecule 6s in polymerized molecule 1.

If the ratio of molecule 6 to molecule 7 in polymerized molecule 1 is kept at 68 : 32 and the molecules 6 and 7 are randomly exchanged, the average ratio of five consecutive molecule 6s is 19%. However, when looking at the calculation cases in which the generation, division, and disappearance of cell-like shapes continued, the ratio of five consecutive cases moved in the direction higher than 19% in all cases. This means that if, during the mutation process, the reaction network that maintains the cell shape fails to maintain the shape due to lack of energy molecules (molecule 25) if the ratio of 5 consecutive is not high, the cell shape region will disappear. Conversely, a phenomenon is expected to be observed in which the cell-like shape that retains the polymerized molecule 1 with a high ratio of 5 contiguous molecule 6s survives.

When the amount of energy resources (molecule 20s) input into the computational lattice space is varied, the more abundant the energy resources, the more frequently the cell-like shapes survive. Furthermore, when the amount of energy resources input is reduced after a certain time step, the cell-like region that retains the polymerized molecule 1 with a high ratio of 5 consecutive survives. Since this region is able to maintain high energy production, it is thought that the cell shape is able to survive and adapt to environments with low energy resources (which would not normally survive). The model shows that even in an environment with limited energy resources, individuals with informants that can generate a lot of energy will survive, and it is thought that the basic process of evolution in response to the environment has been created.

The surviving cell-like regions were found to have a greater ratio of 5 conservative molecular 6s than the random average. This corresponds to spontaneous selection for genetic information (individuals that happen to be in favorable conditions survive). This suggests that it indicates the primitive evolution, in which cells with informants is more favorable than the average remain.

In the Ishida[23]) model, only three types of cell shape emergence, replication, and metabolism were realized, but this model may have shown that the emerged cellular shape also has "evolutionary potential".

5. Conclusion

Based on the Ishida[23]) model, it was constructed a model that explicitly incorporates reactions of energy metabolism by adding three new molecules in addition to the 15 molecules. This model was able to produce cell-like shapes with the ability to evolve to an energy resource environment. With this model, a cell-like shape with the four conditions of a cell, boundary, metabolism, replication, and evolution, could be emergently created by assuming several chemical reactions and molecular polymerization.

This model is currently a simple model, but as the number of molecule type increases and the degree of freedom of the reaction increases, under sufficient energy supply to sustain the reaction, new reaction paths and multiple reaction loops may be formed. This could potentially reproduce the evolution to more complex cells.

If this mechanism can be applied to actual chemical reactions, it may provide a basis for considering the synthesis of minimal cells (protocells) or the self-organized synthesis of nanomachines.

On the other hand, the weakness of this model is that it assumes that the formation and decomposition of 100 linked polymerized molecules can be easily realized. However, this cannot be easily achieved in a natural reaction environment unless a catalyst or other means is already present. There are still difficulties in using this model to explain the emergent process in which the first life simultaneously maintains all four conditions in the absence of catalyst.

To solve this issue, as the next stage of research, it is necessary to construct a computational model that includes the emergence process of long polymerized molecules with catalytic functions, starting from single molecules and light polymerized molecules polymerized by a few molecules (which can polymerize and degrade naturally by the action of minerals or other substances without protein catalysts), and simulate life emergence including "catalyst emergence".

**Conflicts of Interest**

The author declares no conflict of interest regarding the publication of this paper.

**Funding Statement**

This research was supported by grants from Japan Society for the Promotion of Science, KAKENHI Grant Number 19K04896.

**Acknowledgments**

The authors would like to thank Enago (www.enago.jp) for the English language review.